\documentclass[prl,aps,twocolumn,showpacs]{revtex4}
\usepackage{graphicx}
\usepackage{braket}
\usepackage{times,dsfont,amssymb,amsmath,amsthm,amsfonts,amsbsy,mathrsfs,bm,bbm,graphicx,graphics,epsfig,multirow,mathtools,color,bbold}

\begin{document}

\title{What can quantum optics say about computational complexity theory?}

\author{Saleh Rahimi-Keshari, Austin P. Lund, and Timothy C. Ralph}
\affiliation{Centre for Quantum Computation and Communication Technology, \\
	School of Mathematics and Physics, University of Queensland, St Lucia,
	Queensland 4072, Australia}
\date{\today}

\begin{abstract}

Considering the problem of sampling from the output photon-counting probability distribution of a linear-optical network for input Gaussian states, we obtain results that are of interest from both quantum theory and the computational complexity theory point of view. We derive a general formula for calculating the output probabilities, and by considering input thermal states, we show that the output probabilities are proportional to permanents of positive-semidefinite Hermitian matrices. It is believed that approximating permanents of complex matrices in general is a \#P-hard problem. However, we show that these permanents can be approximated with an algorithm in $\text{BPP}^{\text{NP}}$ complexity class, as there exists an efficient classical algorithm for sampling from the output probability distribution. 
We further consider input squeezed-vacuum states and discuss the complexity of sampling from the probability distribution at the output.
\end{abstract}


\maketitle


{\it Introduction.}---Boson Sampling is an intermediate model of quantum computation that seeks to generate random samples from a probability distribution of photon (or, in general, Boson) counting events at the output of an $M$-mode linear-optical network consisting of passive optical elements, for an input with $N$ of the modes containing single photons and the rest in the vacuum states~\cite{AA}. There is great interest in this particular computational problem as this task, despite its simple physical implementation, is strongly believed to be a problem that cannot be efficiently simulated classically. This has led to several proof of principle experiments realizing small-scale Boson Sampling~\cite{BRO13,SPR13,TIL13,CRE13} and investigations of its characterization~\cite{Spagnolo,Tichy} and implementation~\cite{Motes}.
 
In Boson Sampling, the photon-counting probabilities are proportional to the modulus squared of permanents of complex matrices, which in the case of single-photon detections, are submatrices of the unitary matrix describing the linear-optical network~\cite{Scheel}. It has been proved that exactly computing the permanent of matrices is difficult (\#P-hard in complexity theory)~\cite{Valiant,A08}, and it is in a class that contains the polynomial hierarchy of complexity classes~\cite{Toda}. More recently, it was proved that approximating squared permanents of real matrices to within a multiplicative error is also \#P-hard, and it is believed this is the case for modulus-squared permanents of arbitrary complex matrices~\cite{AA}. Based on this key observation, Aaronson and Arkhipov have shown that Boson Sampling cannot be classically simulated unless the polynomial hierarchy collapses to the third level, a situation believed to be highly unlikely.  

In this paper, we consider the problem of sampling from the photon-counting probability distribution at the output of a linear-optical network for input Gaussian states, which is referred to as {\it Gaussian Boson Sampling}. We derive a general formula for the probabilities of detecting single-photons at the output of the network. Using this formula we show that probabilities of single-photon counting for input thermal states are proportional to permanents of positive-semidefinite Hermitian matrices. However, any classical states can be modeled as a statistical mixture of coherent states, and as a result we show that sampling from the output probability distribution can be performed efficiently on a classical computer. Thus, by using Stockmeyer's approximate counting algorithm~\cite{Stockmeyer,AA}, one can approximate permanents of positive-semidefinite Hermitian matrices in the complexity class $\text{BPP}^{\text{NP}}$, which is less computationally complex than \#P-hard. To the best of our knowledge this result was not previously known.

In addition, we consider squeezed-vacuum states as input to a linear-optical network. We find the probabilities of detecting single photons at the output is proportional to the modulus squared of a quantity $O_N$, which is obtained by summing up $(N-1)!!$ complex terms with $N$ being the number of the detected single-photons. It was recently shown that a specific case of this problem is equivalent to a randomized version of the Boson Sampling problem that cannot be efficiently simulated using a classical computer~\cite{RanSam}. This implies that, following the results from~\cite{AA}, at least for this specific problem approximating $|O_N|^2$ is \#P-hard. However, it would be surprising if this problem was the only case of the general problem of Boson Sampling with squeezed-vacuum states, for which approximating $|O_N|^2$ is a \#P-hard problem. Such considerations may help a complexity theorist to identify other \#P-hard problems.

{\it Brief review of previous works.}---If the photons behaved as classical particles, i.e., there were no interferences (the nonclassical effect) between them as they scattered by a linear-optical network, the output probabilities would be permanents of matrices with non-negative elements~\cite{AA}. In this classically simulatable situation, one can use Stockmeyer's approximate counting algorithm~\cite{Stockmeyer} to approximate one particular output probability, even if it is exponentially small, to within a multiplicative error in $\text{BPP}^{\text{NP}}$ (in the third level of the polynomial hierarchy); for a short description of this algorithm see the supplementary information of Ref.~\cite{RanSam} or theorem 4.1 of Ref.~\cite{AA}.  This algorithm was further improved and it was shown that the approximation can be done in BPP (in the second level of the polynomial hierarchy)~\cite{Jer2004}. The probability $p$ is approximated with $\tilde p$ to within a multiplicative factor of $g$, if $p/g\leq \tilde{p} \leq g p$ for $g\geq 1+1/h(N)$, where $h(N)$ is a polynomial function in the size of the problem $N$ (number of detected single photons). Throughout this paper we refer to this form of approximation only.

Aaronson and Arkhipov~\cite{AA} have shown that if there is a polynomial-time classical algorithm for Boson Sampling with single-photon inputs, then one could use Stockmeyer's approximate counting algorithm to approximate the probability of detecting a particular configuration of output photons in $\text{BPP}^{\text{NP}}$. This would then approximate the modulus squared of the permanent of a submatrix of a unitary matrix. However, on the other hand it was shown that this approximation is \#P-hard~\cite{AA}, as the elements of a unitary matrix are, in general, complex numbers, and an algorithm for this problem can solve all of the problems in the entire polynomial hierarchy~\cite{Toda}. Therefore, the polynomial hierarchy of complexity classes would collapse to the third level, 
if there exists a classical algorithm that can efficiently simulate Boson Sampling, a highly implausible situation~\cite{AA}. It was also shown in Ref.~\cite{AA} that, modulo two conjectures, even sampling from a probability distribution that is an approximation of the output probability distribution is classically intractable as well. This form of sampling is referred to as the {\it approximate} Boson Sampling as opposed to the {\it exact} Boson Sampling that is for sampling from the exact output probability distribution. Here we consider exact Boson Sampling only.

\begin{figure}[tpb]
\centering
	\includegraphics[width=1\columnwidth]{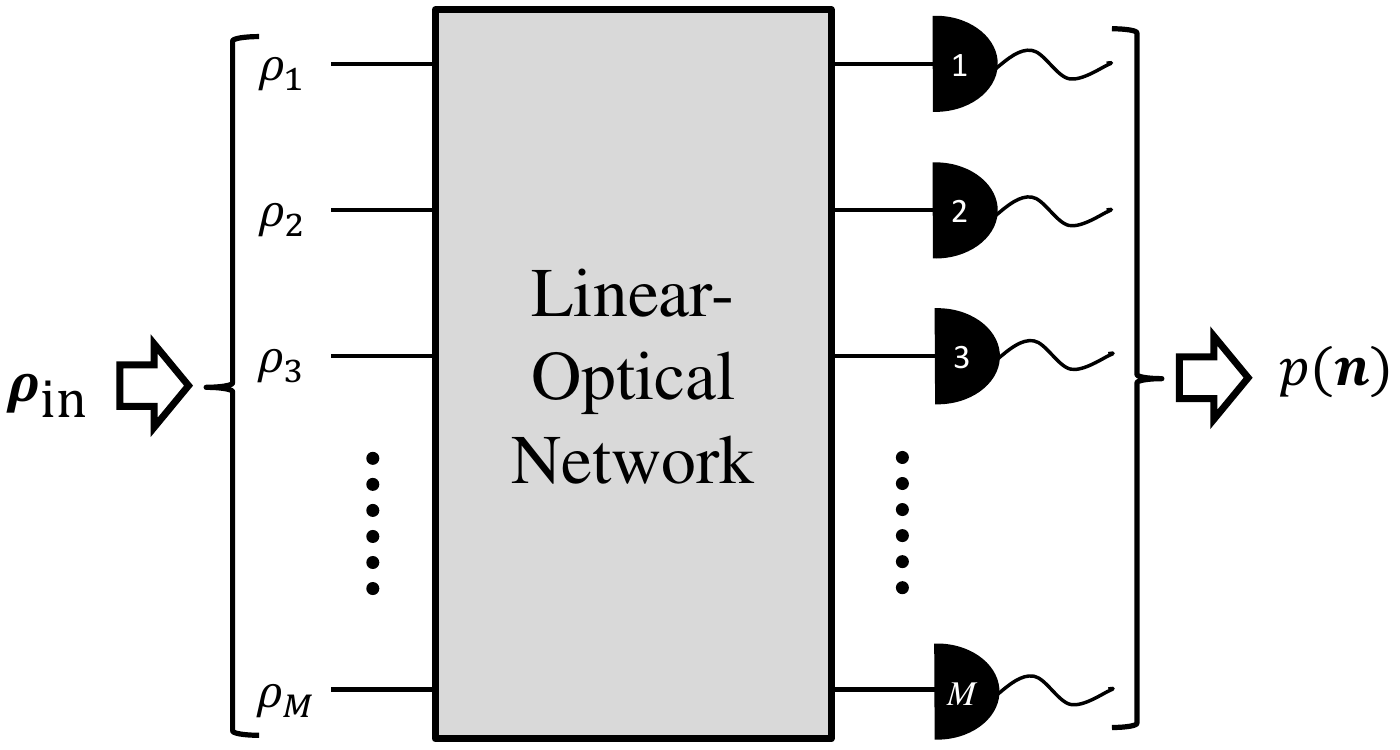}
	\caption{In the Gaussian Boson Sampling problem for a given product Gaussian input state, $\bm\rho_{\text{in}}=\otimes_{s=1}^M \rho_s$, and a unitary matrix describing the network, one samples from the output probability distribution $p(\mathbf{n})$. }
	\label{Gen-Bos-Sam}
\end{figure}

{\it Photon-counting probability distribution.}---In the Gaussian Boson Sampling problem, we consider the photon-counting probability distribution at the output of an $M$-mode linear-optical network for an input multimode Gaussian quantum state $\bm\rho_{\text{in}}$, which is a product state of the individual states $\{\rho_s\}$ in each mode; see Figure~\ref{Gen-Bos-Sam}. 
We are then interested in the output probabilities of detecting $N$ single photons,
\begin{equation}
\label{prob-dist}
p(\mathbf{n})=\text{Tr}[\bm\rho_{\text{out}}\ket{\mathbf{n}}\bra{\mathbf{n}}],
\end{equation}
where $\mathbf{n}=(n_1,n_2,n_3,\dots,n_{M})$, $n_s\in\{0,1\}$, $\sum_s n_s=N$, and  $\bm\rho_{\text{out}}=\mathcal{U}\bm\rho_{\text{in}}\mathcal{U}^{\dagger}$ with $\mathcal{U}$ being the unitary operator that describes the linear-optical network. In practice, one must use photon-number-resolving detectors in order to distinguish the single-photon events from events in which a detector registers more than one photon. Hence, in Gaussian Boson Sampling, inefficiency of detectors will cause errors in distinguishing the events. Note, however, that the errors can be minimized if the mean-photon number at the input is much less than the number of modes. Also, for the exact Boson Sampling case, the detection probabilities are allowed to be exponentially small.

A linear-optical network can also be uniquely represented by an $M{\times}M$ unitary matrix $\mathbf{U}$ that relates the creation operators of the output modes $\hat{b}_k^{\dagger}$ to those of the input modes $\hat{a}_j^{\dagger}$,
\begin{equation}
\hat{b}_j^{\dagger}=\mathcal{U}\hat{a}_j^{\dagger}\mathcal{U}^{\dagger}=\sum_{k=1}^{M} \mathbf{U}_{jk} \hat{a}_k^{\dagger}.
\label{U}
\end{equation}

For a multimode input coherent state $\ket{\bm{\alpha}}$, where $\bm\alpha=(\alpha_1,\alpha_2,\alpha_3,\dots,\alpha_{M})$, the output state is also a multimode coherent state. By using the relation~(\ref{U}), we have
\begin{align}
\mathcal{U}\ket{\bm\alpha}\!=\!\prod_{j=1}^{M}\!D(\mathcal{U}\hat{a}_j^{\dagger}\mathcal{U}^{\dagger},\alpha_j)\ket{0}
=\!\prod_{k=1}^{M}\!D(\hat{a}_k^{\dagger},\beta_k)\ket{0}=\!\ket{\bm{\beta}},\nonumber
\end{align}
where $D(\hat{a}_j^{\dagger},\alpha_j)=\exp(\alpha_j\hat{a}^{\dagger}_j-\bar\alpha_j\hat{a}_j)$ is the displacement operator for mode $\hat{a}_j$ with $\bar{\alpha}_j$ being the complex conjugate of $\alpha_j$, and the output amplitudes are 
\begin{equation}
\beta_k=\sum_{j}^{M}\alpha_j\mathbf{U}_{jk}.
\label{out-amp}
\end{equation}
Using this equation the probability distribution~(\ref{prob-dist}) is then given by
\begin{equation}
p(\mathbf{n})=e^{-I}\prod_{k=1}^{M}|\beta_k|^{2n_k},
\label{prob-coh}
\end{equation}
where  $I=\sum_{k}^{M}|\beta_k|^2=\sum_{j}^{M}|\alpha_j|^2$. This probability distribution can be efficiently calculated using a classical computer. This implies that there exists an efficient classical algorithm for  Boson Sampling with coherent states. Note, however, that coherent states are useful for efficiently characterizing linear-optical networks that are indispensable for the classical verification of Boson Sampling in practice~\cite{charact}.

In deriving a general formula for calculating the probability distribution~(\ref{prob-dist}), without loss of generality, we make two assumptions about input Gaussian states for Gaussian Boson Sampling. First, we assume that the input states have zero first order moments. This is because any displacement operations before the linear-optical network are equivalent to some displacement operations at the output, which will not change the correlations between output states~\cite{Zhang}. Second, we assume the covariance matrices of the Gaussian states $\rho_s$ are diagonal with the variance in the $x$ quadrature, $V_{x_s}$, being larger than or equal to the variance in the $p$ quadrature, $V_{p_s}$. The reason is that, in general, any local phase-shift operation before the linear-optical network can be absorbed into the unitary operation describing the network. We use the $Q$ function to represent each input Gaussian state $\rho_s$
\begin{align}
Q_s(\alpha_s)
=\frac{\sqrt{\mu_s^2-4\lambda_s^2}}{\pi} \exp\left[\lambda_s(\alpha_s^2+\bar{\alpha}_s^2)-\mu_s |\alpha_s|^2 \right],
\label{Qin}
\end{align}
where
\begin{equation}
\lambda_s=\frac{1}{2V_{p_s}+2}-\frac{1}{2V_{x_s}+2}, \ \ \ \ 
\mu_s=\frac{1}{V_{x_s}+1}+\frac{1}{V_{p_s}+1}, \nonumber
\end{equation}
and for the vacuum state $V_x=V_p=1$. The parameter $\lambda_s$ is between zero (when $V_{p_s}=V_{x_s}$) and infinity (for infinite squeezing), and $\mu_s$ is between zero (for infinite variances) and one (for pure states). The $Q$ function of the output state using Eq.~(\ref{out-amp}) can be calculated as
\begin{align}
Q_{\text{out}}(\bm{\alpha})&=\frac{1}{\pi^M}\bra{\bm\alpha}\mathcal{U}\bm\rho_{\text{in}}\mathcal{U}^{\dagger}\ket{\bm\alpha}
=\frac{1}{\pi^M}\bra{\bm\eta}\bm\rho_{\text{in}}\ket{\bm\eta}\nonumber \\
&=\prod_{s=1}^{M} Q_{s}\left(\sum_{j=1}^{M} \alpha_j \bar{\mathbf{U}}_{js}\right).
\end{align}
where $\ket{\bm\eta}=\mathcal{U}^{\dagger}\ket{\bm\alpha}=\ket{\bm{\alpha} \bar{\mathbf{U}}}$ is an $M$-mode coherent state. By using the expression for the input $Q$ function (\ref{Qin}), the output $Q$ function can be written in this compact form
\begin{equation}
\label{Qout}
Q_{\text{out}}(\bm{\alpha})=\frac{K}{\pi^M} \exp\left[\vec{\bm{\alpha}} \begin{pmatrix}-\mathbf{D} & \mathbf{C}\\ \bar{\mathbf{C}} & 0\end{pmatrix} \vec{\bm{\alpha}}^{\dagger}\right],
\end{equation}
with $\vec{\bm\alpha}{:=}(\!\alpha_1,\dots,\alpha_M,\bar{\alpha}_1,\dots,\bar{\alpha}_M\!)$,
 $K{=}\prod_{s=1}^{M}\sqrt{\mu_s^2-\!4\lambda_s^2}$, $\mathbf{C}{=}\mathbf{U}\bm{\lambda}\mathbf{U}^T$, $\mathbf{D}{=}\mathbf{U}\bm{\mu}\mathbf{U}^{\dagger}$,
where $\bm{\lambda}{=}\text{diag}(\lambda_1,\dots,\lambda_M)$ and $\bm{\mu}{=}\text{diag}(\mu_1,\dots,\mu_M)$.
Now by using this $Q$ function, the probability distribution~(\ref{prob-dist}) is then given by
\begin{equation}
p(\mathbf{n})=(\pi)^M\int_{\mathds{C}^{M}} d^{2M}\!\bm{\alpha} Q_{\text{out}}(\bm{\alpha}) P_{\mbox{\scriptsize\boldmath$n$}\mbox{\scriptsize\boldmath$n$}}(\bm{\alpha}),
\end{equation}
where
\begin{equation}
P_{\mbox{\scriptsize\boldmath$n$}\mbox{\scriptsize\boldmath$n$}}(\bm{\alpha})=\prod_{s=1}^{M} e^{|\alpha_{s}|^2}
\partial_{\alpha_{s}}^{n_{s}}
\partial_{\bar{\alpha}_{s}}^{n_{s}}
\delta^{2}(\alpha_{s})
\label{P-func}
\end{equation}
is the $P$ function of the number state $\ket{\mathbf{n}}\bra{\mathbf{n}}$, $n_s\in\{0,1\}$, with $\partial_{\alpha}^{n}\coloneqq\partial^{n}/\partial\alpha^n$ and $\delta^2(\alpha)\equiv\delta\bigl(\mathrm{Re}(\alpha)\bigr) \delta\bigl(\mathrm{Im}(\alpha)\bigr)$ ~\cite{MandelWolf}. Integration by parts yields
\begin{equation}
p(\mathbf{n})=K\prod_{s=1}^{M} \partial_{\alpha_{s}}^{n_{s}} \partial_{\bar{\alpha_{s}}}^{n_{s}}
 e^{F(\bm{\alpha},\bar{\bm{\alpha}})}\bigg|_{\alpha_s=0},
 \label{prob-N-0}
\end{equation}
where
\begin{equation}
F(\bm{\alpha},\bar{\bm{\alpha}})=\vec{\bm{\alpha}}
\begin{pmatrix}
\tilde{\mathbf{D}} & \mathbf{C}\\
\bar{\mathbf{C}} & 0
\end{pmatrix}
\vec{\bm{\alpha}}^{\dagger},
\label{func}
\end{equation}
with $\tilde{\mathbf{D}}=\mathbb{1}-\mathbf{D}$, $\mathbb{1}$ being the $M\times M$ identity matrix. In the above expression, we have to take $2N$ derivatives with respect to independent variables $\{\alpha_s,\bar{\alpha}_s|n_s\neq 0\}$ at $\bm\alpha=0$; hence, that expression can be written as
\begin{equation}
p(\mathbf{n})=K
\sum_{r=1}^{\infty} L(2N;F,r),
 \label{prob-N-1}
\end{equation}
where $L(2N;F,r)$, analogous to distributing distinguishable balls into indistinguishable boxes, can be understood as a sum over all possible ways to distribute $2N$ derivatives (balls) among $r$ functions (boxes), $\partial^{i_1}F,\dots,\partial^{i_r}F$, such that $\sum_{s=1}^{r} i_s=2N$ and $i_s\neq 0$. As $F(\bm{\alpha},\bar{\bm{\alpha}})$ is a second order polynomial in $\bm\alpha$ and $\bar{\bm\alpha}$, and $\partial^{i_{s}}F|_{\bm\alpha=0}=0$ for $i_s\neq2$, only $L(2N;F,N)$ for $i_s=2$ is nonzero. Therefore, we obtain the desired formula for calculating the probabilities of $N$ single-photon detections as
\begin{equation}
p(\mathbf{n})=K \sum_{i}^{(2N-1)!!} \prod_{l=1}^{N} \frac{\partial^2 F}{\partial {X}^i_{2l-1} \partial {X}^i_{2l}},
\label{Pn}
\end{equation}
where the sum is over $(2N-1)!!$ possible ways of distributing $2N$ balls ($\partial/\partial{X^{i}_{l}}$ where $\{X^{i}_{l}\}_{l=1}^{2N}=\{\alpha_s,\bar{\alpha}_s|n_s\neq 0\}$) into $N$ boxes ($F$'s) such that each box contains two balls. In the following, by using this new formula, we consider two cases of thermal states and squeezed-vacuum states as inputs.

{\it Boson Sampling with thermal states.}---If one subjects $M$ thermal states with the same temperatures, i.e., $\mu_s=2/(V_s+1)=\mu$ and $\lambda_s=0$ for all $s$, to a linear-optical network, we have $\mathbf{D}=\mu\mathbb{1}$ and $\mathbf{C}=0$ in the output $Q$ function (\ref{Qout}). In this case the output $Q$ function is identical to the input $Q$ function and no correlation is created. Here we assume the input thermal states have different temperatures such that the matrix $\mathbf{D}$ is not diagonal, in general. In this case, the formula (\ref{Pn}) becomes
\begin{equation}
p(\mathbf{n})=\left( \prod_{s=1}^{M}\mu_s \right) \sum_{i}^{N!} \prod_{l=1}^{N} \frac{\partial^2 }{\partial {X}^i_{2l-1} \partial {X}^i_{2l}}\left[{\bm{\alpha}}\tilde{\mathbf{D}}{\bar{\bm{\alpha}}}^T\right] ,
\label{th-per}
\end{equation}
where $\{X^{i}_{2l-1}\}_{l=1}^{2N}{=}\{\alpha_s|n_s{=}1\}$ and $\{X^{i}_{2l}\}_{l=1}^{2N}{=} \{\bar{\alpha}_s|n_s{=}1\}$. By comparing this equation with the definition of permanent~\cite{AA}, it can be seen by inspection that
\begin{equation}
p(\mathbf{n})=\left( \prod_{s=1}^{M}\mu_s \right) \text{Per}\left([\tilde{\mathbf{D}}]_{N\times N}\right).
\label{per-th}
\end{equation}
Thus, the probabilities of having $N$ simultaneous single-photon detections at the output are proportional to permanents of $N\times N$ submatrices of the Hermitian matrix $\tilde{\mathbf{D}}$, denoted by $[\tilde{\mathbf{D}}]_{N\times N}$. The submatrices are obtained by removing $M-N$ rows and the same $M-N$ columns corresponding to those output modes from which no photon was detected. Notice that we have $\tilde{\mathbf{D}}=\mathbf{U}\tilde{\bm\mu}\mathbf{U}^{\dagger}$, where the elements of matrix $\tilde{\bm\mu}$ are $(1-\mu_j)\delta_{ij}\geq 0$; hence, $\tilde{\mathbf{D}}$ and its principal submatrices $[\tilde{\mathbf{D}}]_{N\times N}$ are positive-semidefinite Hermitian matrices.

We now see whether Boson Sampling with thermal states can be efficiently simulated classically. Each input thermal state can be expressed as a Gaussian statistical mixture of coherent states due to the Glauber-Sudarshan representation~\cite{Glauber,Sudarshan}
\begin{equation}
\rho^{\text{th}}_j=\int_{\mathds{C}}\mathrm{d}^{2}\!{\alpha_j} P^{\text{th}}_j(\alpha_j) \ket{\alpha_j}\bra{\alpha_j},
\label{G-S-rep}
\end{equation}
where $P^{\text{th}}_j(\alpha_j)$ is a Gaussian $P$ function for the thermal state to input mode $j$. By choosing a random set of input coherent states with amplitudes $\{\alpha_j\}_{j=1}^{M}$ from the probability distributions $\{P^{\text{th}}_j(\alpha_j)\}_{j=1}^{M}$, one can efficiently find the amplitudes of output coherent states $\{\beta_k\}_{k=1}^{M}$ and the probability distribution from Eq.~(\ref{prob-coh}). This implies that there exists an efficient classical algorithm for Boson Sampling with thermal states. Hence, using Stockmeyer's approximate counting algorithm~\cite{Stockmeyer}, the probability (\ref{per-th}) for a specific $\mathbf{n}$ can be approximated in $\text{BPP}^{\text{NP}}$. As any arbitrary positive-semidefinite Hermitian matrix $\tilde{\mathbf{D}}^{\prime}$ can be written as 
$\tilde{\mathbf{D}}^{\prime}=\mathbf{U}q\tilde{\bm\mu}\mathbf{U}^{\dagger}$ with $q\geq 1$, we then have $\text{Per}([\tilde{\mathbf{D}}^{\prime}]_{N\times N})=q^N\text{Per}([\tilde{\mathbf{D}}]_{N\times N})$, which is proportional to the output probability~(\ref{per-th}). Therefore, using Stockmeyer's algorithm, the permanent of any arbitrary positive-semidefinite Hermitian matrix, despite having complex number elements, can be approximated in $\text{BPP}^{\text{NP}}$, which is in the third level of the polynomial hierarchy. Unless the polynomial hierarchy collapses to this level, this problem is not \#P-hard.

Based on the above argument, Boson Sampling with any classical input states, i.e., quantum states with non-negative $P$ functions, can be efficiently simulated with a classical computer as well. Notice that the output probabilities can be also calculated by using the output probabilities for input coherent state (\ref{prob-coh}) and the $P$ functions of the input states
\begin{equation}
p(\mathbf{n})=\int_{\mathds{C}^{M}}\hspace{-1em} d^{2M}\!\bm{\alpha} \prod_{k=1}^{M} P_k(\alpha_k)e^{-|\alpha_k|^2} \bigg| \sum_{j}^{M}\alpha_j U_{jk} \bigg|^{2n_k}\!.
\end{equation}
Therefore, according to the above argument, for all of the $P$ functions that are valid probability density functions, the above integral can be approximated in $\text{BPP}^{\text{NP}}$.


{\it Boson Sampling with squeezed-vacuum states.}---Let us now consider squeezed-vacuum states whose variances in the $x$ and $p$ quadratures are $V_{x_s}=e^{2r_s}$ and $V_{p_s}=e^{-2r_s}$, respectively, where $r_s$ is the squeezing parameter for input mode $s$. In this case, we have $\mu_s=1$ for all $s$, $\tilde{\mathbf{D}}=0$, $\lambda_s=(\tanh{r_s})/2$ and $K=\prod_{s=1}^{M}(\cosh{r_s})^{-1}$. Note that if the input states have the same squeezing parameter, $\bm\lambda=
\lambda \mathbb{1}$, $\bm\mu=
\mu\mathbb{1}$ and $\mathbf{U}$ is an orthogonal matrix, then we have $\mathbf{C}= \bm\lambda$ and $\mathbf{D}= \bm\mu$; hence, in this case, according to Eq.~(\ref{Qout}) the output state $\bm\rho_{\text{out}}$ is identical to the input state $\bm\rho_{\text{in}}$ and no correlation is generated.

As the function (\ref{func}) becomes $F(\bm{\alpha},\bar{\bm{\alpha}})=F_1(\bm\alpha)+F_1(\bar{\bm\alpha})$, $F_1(\bm\alpha)=\bm\alpha \mathbf{C} \bm\alpha^T$, we have $\partial_{\alpha_{j}} \partial_{\bar\alpha_{j}}F|_{\bm\alpha=0}=0$, for any $i$ and $j$. Thus, by using the formula~(\ref{Pn}) the probability distribution for detecting $N$ single photons at the output is given by
\begin{align}
p(\mathbf{n})
=\left(\prod_{s=1}^{M}\frac{1}{\cosh{r_s}}\right) \left| \sum_{i}^{(N-1)!!} \prod_{l=1}^{N/2} \frac{\partial^2 F_1(\bm\alpha)}{\partial {X}^i_{2l-1} \partial {X}^i_{2l}}\right|^2,
\label{Pnn-sq}
\end{align}
where $\{X^{i}_{l}\}_{l=1}^{N}=\{\alpha_s|n_s=1\}$. One can immediately see from this distribution that, independent of what the linear-optical network is, the probability of detecting an odd number of single photons at the output is always zero as expected from squeezed-vacuum inputs. The probabilities (\ref{Pnn-sq}) are proportional to the modulus squared of this quantity
\begin{equation}
O_N=\sum_{i}^{(N-1)!!} \prod_{l=1}^{N/2} \frac{\partial^2 F_1(\bm\alpha)}{\partial {X}^i_{2l-1} \partial {X}^i_{2l}},
\end{equation}
which depends on the off-diagonal elements of the matrix $\mathbf{C}$ and the number of detected single photons. Notice that quantity $O_N$ is not a permanent, but it is a sum of $(N-1)!!$ complex numbers. Considering that the matrix $\mathbf{C}$ is symmetric, $c_{ij}=c_{ji}$, we have $\partial_{\alpha_i}\partial_{\alpha_j} F_1(\bm\alpha)=2c_{ij}$, with $i\neq j$. Hence, the above quantity can be written as
\begin{align}
O_N&{=}\!\!\sum_{i_1\neq i_2}(c_{i_{1}i_{2}}\sum_{i_{3}\neq i_4}(c_{i_{3}i_{4}}\!\dots\hspace{-1.5em}\sum_{i_{2k-1}\neq i_{2k}}\hspace{-1em}(c_{i_{2k-1}i_{2k}}\!\dots c_{i_{N-1}i_{N}})\!\dots\!))\nonumber\\
&\times 2^{N/2},
\end{align}
where $i_{1}=1$, $i_l\neq i_1,\dots,i_{l-1}$ for $2 \leq l \leq N$.

For a particular case of Boson Sampling with squeezed-vacuum states, it has been shown that sampling cannot be simulated classically~\cite{RanSam}. Consider an $M$-mode linear-optical network, which consists of $M/2$ beam splitters with a $\pi/2$-phase shifter at one of the input ports and an $M/2$-mode linear-optical network that acts only on half of the output modes of the beam splitters. By feeding this $M$-mode network with $M$ squeezed-vacuum states, the beam splitters generate $M/2$ two-mode entangled (two-mode squeezed-vacuum) states. Then, conditional on detecting $N/2$ single-photons from one particular configuration of the output modes of beam splitters, $N/2$ single-photons in the corresponding other modes are subjected to the $M/2$-mode network, and the problem reduces to that of the original Boson Sampling. This implies that sampling from the single-photon-counting probability distribution at the output of the $M$-mode network cannot be simulated classically, and thus, following the Aaronson and Arkhipov results~\cite{AA}, for at least this type of configuration approximating $|O_N|^2$ is a \#P-hard problem.
It would be surprising if this were the only configuration for which approximating $|O_N|^2$ was \#P-hard, as the squeezed-vacuum states are highly non-classical with a highly singular $P$ function and the output is almost always an entangled state~\cite{Zhang}. This result may be of interest to computational complexity theory as a way of identifying other classically hard problems besides the computing of permanents.

{\it Conclusion.}---We have presented new results that are interesting from quantum computation, computational complexity theory, and optics perspectives, by
considering the problem of sampling from the output
probability distribution of a linear-optical network for input Gaussian states. Our results show that the consideration of problems in quantum optics can help to classify and identify new problems in computational complexity theory. There are two interesting open questions. The first question is whether permanents of positive-semidefinite Hermitian matrices can be approximated with an algorithm similar to the algorithm for matrices with non-negative entries~\cite{Jer2004} in BPP. Note that the probabilities (\ref{per-th}) for input thermal states and (\ref{Pnn-sq}) for squeezed-vacuum states are special cases of the formula (\ref{Pn}) for general squeezed thermal input states. By adding sufficient thermal noise to input squeezed-vacuum states, they will become classical with positive $P$ function and as shown, sampling can be simulated classically. Hence, the second question is, as we add thermal noise to pure squeezed-vacuum input states, at what point does sampling become classically simulatable; does entanglement play any role?

{\it Acknowledgement.}---We thank Howard Wiseman for the discussions and Scott Aaronson and Alex Arkhipov for their comments. This research was conducted by the Australian Research Council Centre of Excellence for Quantum Computation and Communication Technology (Project number CE110001027).

\end{document}